\documentclass[aip,jap,superscriptaddress,floatfix,reprint]{revtex4-1}

\usepackage{graphicx,amssymb,bm,natbib,amsfonts,amsmath,graphics,color}
\usepackage[bookmarks=false,pdfstartview=FitH]{hyperref}

\newcommand {\degree}{$^{\circ}$}

\begin{document}

\title{Vacuum space charge effect in laser-based solid-state photoemission spectroscopy}

\author{J. Graf} 
\affiliation{Materials Sciences Division, Lawrence Berkeley National Laboratory, Berkeley, CA 94720, USA}
\affiliation{Department of Physics, University of California, Berkeley, CA 94720, USA}
\author{S. Hellmann}
\affiliation{Institute for Experimental and Applied Physics, University of Kiel, D-24098 Kiel, Germany}
\author {C. Jozwiak}
\affiliation{Materials Sciences Division, Lawrence Berkeley National Laboratory, Berkeley, CA 94720, USA}
\affiliation{Department of Physics, University of California, Berkeley, CA 94720, USA}
\author{C.L. Smallwood}
\affiliation{Department of Physics, University of California, Berkeley, CA 94720, USA} 
\author{Z. Hussain}
\affiliation{Advanced Light Source, Lawrence Berkeley National Laboratory, Berkeley, California 94720, USA}
\author{R. A. Kaindl}
\affiliation{Materials Sciences Division, Lawrence Berkeley National Laboratory, Berkeley, CA 94720, USA}
\author{L. Kipp}
\affiliation{Institute for Experimental and Applied Physics, University of Kiel, D-24098 Kiel, Germany}
\author{K. Rossnagel}
\affiliation{Institute for Experimental and Applied Physics, University of Kiel, D-24098 Kiel, Germany}
\author{A. Lanzara}\email{ALanzara@lbl.gov}
\affiliation{Materials Sciences Division, Lawrence Berkeley National Laboratory, Berkeley, CA 94720, USA} \affiliation{Department of Physics, University of California, Berkeley, CA 94720, USA}
\date {\today}
\begin {abstract}
We report a systematic measurement of the space charge effect observed in the few-ps laser pulse regime in laser-based solid-state photoemission spectroscopy experiments. The broadening and the shift of a gold Fermi edge as a function of spot size, laser power, and emission angle are characterized for pulse lengths of 6 ps and 6 eV photon energy. The results are used as a benchmark for an $N$-body numerical simulation and are compared to different regimes used in photoemission spectroscopy. These results provide an important reference for the design of time- and angle-resolved photoemission spectroscopy setups and next-generation light sources.
\end {abstract}
\keywords{Space charge effects, Laser based ARPES}
\maketitle

\section{Introduction}
\label{sec:INTRODUCTION}

Electrons traveling in vacuum to a measurement apparatus will repel each other because of their mutual Coulomb interaction. If a cloud of electrons is sufficiently dense, this vacuum space charge effect can alter the time, energy, and spatial spread of the cloud, limiting experimental resolution and resulting in systematic measurement errors.\cite{Clauberg89,Farkas90,Gilton90,Girardeau91}

The space charge effect has been addressed mostly for the design of electron guns for time-resolved electron diffraction experiments \cite{Aeschlimann95,Qian02,Farkas90} and has been discussed in terms of $N$-body numerical simulations, mean-field models, and fluid dynamics. 
Recently the space charge effect has also been discussed in angle-resolved photoemission spectroscopy (ARPES) experiments \cite{Hellmann09,Passlack06,Zhou05b}, where it has been shown to depend primarily on four parameters: photon energy, pulse duration, incident beam spot size, and the number of electrons photo-emitted per pulse.  These previous studies explored two different regimes: i) the relatively high photon energy ($\geq 35$ eV) and long pulse duration ($\approx 60$ ps)\cite{Zhou05b} regime, and ii) the very low photon energy (two-photon photoemission with 3 eV photons) and short pulse duration ($\approx 40$ fs) regime.\cite{Passlack06} 
The shape of the electron cloud is very different in these two regimes. In the first case, the wide distribution of electron kinetic energy and long pulse duration produce a cloud extended mainly along the direction normal to the sample surface, while in the second case the narrow electron kinetic energy distribution and short pulse duration produce a disk-shaped electron cloud parallel to the sample surface. The former case of a more three-dimensional cloud requires numerical simulations while a disk-shaped 
cloud can be characterized analytically using a simple power law.\cite{Zhou05b,Passlack06,Hellmann09}

With the recent advent of laser-based photoemission experiments, the need to extend these studies to an intermediate regime becomes imperative. A recent study of the space charge effect using laser based ARPES with 7 eV photon energy and 10 ps long pulse suggests that the space charge effect plays a minor role in this regime thanks to the mirror charge effect that nearly cancels the space charge effect in the 1 to 10 ps pulse duration.\cite{Liu08}

Here we expand these results by reporting a systematic study of the space charge effect on polycrystalline gold in laser-based photoemission spectroscopy in the intermediate regime of moderate pulse lengths ($\approx$ 6 ps) and low photon energy (6 eV). We find in the limit of small spot size ($\approx$ 20 $\mu$m diameter) a shift of the Fermi edge on the order of 2 meV and a broadening larger than 10 meV starting already at 100 electrons per pulse. More generally, we find that a single power law cannot describe the broadening or the Fermi edge shift as a function of electrons per pulse and spot size in the various ARPES regimes. To explore the non-trivial relation between the various experimental parameters, we modeled our data with a molecular-dynamics model which employs full $N$-body numerical simulations of the electron propagation from the surface to the detector.\cite{Hellmann09} This model is able to reproduce quantitatively the space charge effect seen in the long and short pulse regimes as well as in the intermediate regime explored here. The large range of applicability of this model allows one to identify the functional dependence of the spectral shift and broadening on the number of electrons per pulse and spot size. We found that one of the key parameters affecting the different power laws observed in different ARPES regimes is the maximum kinetic energy of the initial photo-electron distribution. We believe that this study will be crucial for guiding next generation high resolution setups.

\section{Experimental technique}
Experiments were performed on a polycrystalline gold sample scratched in air and inserted through a load lock into a UHV analysis chamber at low temperature (15-20 K). Pressure remained below $5 \times 10^{-11}$ Torr throughout the experiment.
In Fig.\ 1 we show a diagram of the experimental setup. The sample geometry is detailed in the inset.  The spot size was characterized by using a wire mesh mounted on the same sample holder as the gold sample.

\begin{figure} \includegraphics[width=8cm]{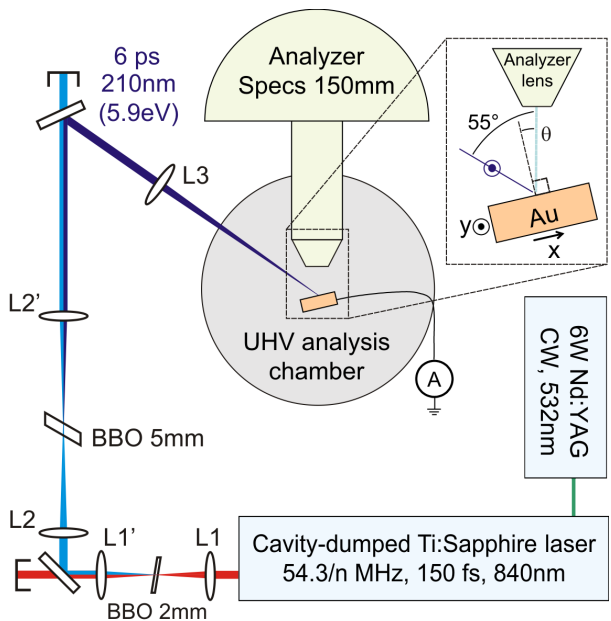}
\caption{\label{fig:1}(Color online) Laser-ARPES experimental setup. The focal lengths are: L1=10cm, L2=20cm and L3=30cm.}\end{figure}

The laser system is a cavity-dumped, mode-locked Ti:Sapphire oscillator (Coherent Mira), pumped with a 6 W frequency-doubled Nd:YVO$_4$ laser. The oscillator generates $\approx$ 150 fs pulses at 840 nm at a repetition rate of 54.3/$n$ MHz ($n=1,2,3, \dots$) with pulse energies as high as 100 nJ. The photon energy of the laser beam was frequency-quadrupled via cascaded, type-I phase-matched second harmonic generation in two beta barium borate (BBO) crystals of 2 mm and 5 mm thickness. The resulting 210 nm (5.9 eV) pulse was estimated to be $\approx$ 6 ps long due mainly to group velocity mismatch. The UHV chamber window is made of UV- grade fused silica. We adjusted the pulse energy by tuning the power applied to the cavity dumper Bragg cell. The focusing lens (L3 in Fig.\ 1) has a 300 mm focal length. We varied the spot size by defocussing L3.

Assuming a work function of about 4.4 eV, electrons at the Fermi energy exit the sample with 1.5 eV kinetic energy, corresponding to a velocity of 0.7 $\mu$m/ps. With a pulse lasting 6 ps, the initial electron spread normal to the surface is thus about 4 $\mu$m. With a spot size diameter of about 20-100 $\mu$m, the initial electron cloud is a flat cylinder. This is an intermediate case between the fs laser-based data and the synchrotron-based  data.\cite{Zhou05b,Passlack06}

The integrated photoemission signal was measured using a Phoibos 150 mm hemispherical analyzer from Specs with a 0.2 mm entrance slit and 4 eV pass energy, yielding an energy resolution of 4 meV\@. The sample was electrically grounded through a picoammeter, which allowed the measurement of the total photo-electron yield. All the spectra were averaged over $\approx 18$\degree\ along the analyzer slit except where explicitly stated otherwise. The angle between the analyzer and the laser beam is fixed at 55\degree\ and the light incident on the sample is $s$-polarized.

\begin{figure}\includegraphics[width=8cm]{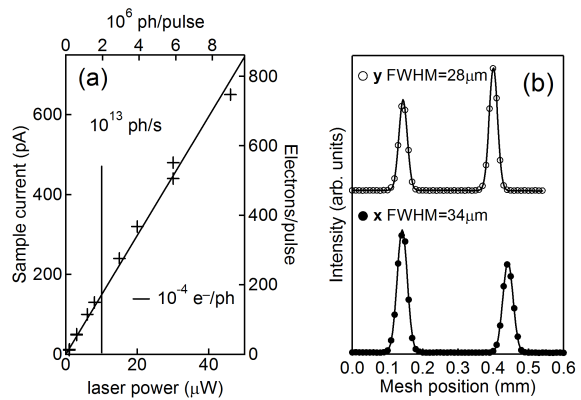}
\caption{\label{fig:2}(Color online) Spot size measurements. (a) Photo-electron yield as a function of laser power. (b) Typical \textbf{x} and \textbf{y} scan of the integrated photoemission intensity from the W mesh. The \textbf{y} scan is vertically shifted}\end{figure}

Fig.\ 2 shows the characterization of the pulse energy and spot size. The total photo-electron yield as a function of the laser power is shown in Fig.\ 2(a). The repetition rate was set to 5.4 MHz. The solid line is a linear fit giving a yield of $1\times 10^{-4}$  e$^-$/photon. For comparison, the typical yield for a photon energy of 35 eV measured at BL 10.0.1 at the Advanced Light Source (ALS) is about 0.2 e$^-$/photon ($1\times 10^{12}$ photons/s with $\approx 300$ nA yield).\cite{Zhou05b} The difference can be mainly attributed to the range of electronic states probed: 6 eV photons only excite states in the Au 5d-6s valence band with binding energy less than about 1.5 eV, whereas 35 eV photons can excite the entire valence band.\cite{Zhou05b} Additionally, in the latter case secondary electrons amplify the photo-electron yield noticeably.\cite{Hufner}

The spot size was measured by scanning a 100$\times$100 lines/inch tungsten mesh at the sample position while measuring the integrated photoemission intensity. The mesh wire diameter was 25 $\mu$m. The $\theta$ angle (see Fig.\ 1) was set to 55\degree\ during the mesh scan so that the \textbf{x} and \textbf{y} scanning directions were perpendicular to the laser beam. Fig.\ 2(b) shows a typical scan along the \textbf{x} and \textbf{y} direction obtained from this procedure. We then estimated the spot size by deconvolving the peak's full width at half maximum (FWHM) using the known mesh wire diameter. The reported diameter is the square root of the \textbf{x} and \textbf{y} FWHM product, corresponding to the diameter of a disc of equal area. The ratio between the FWHM along \textbf{x} and \textbf{y} is not constant but was smaller than 3. We note that with an incident Gaussian beam of 2 mm diameter, the diffraction limited spot size with a 300 mm focal length lens is about 15 $\mu$m. For the FWHM along the \textbf{x} direction, we used a corrected wire diameter to account for the fact that only the part of the wire facing the analyzer could contribute to the measured signal. It can be shown that the effective wire diameter in this case is $25\times (1+\cos(55^\circ))/2=20$ $\mu$m. The small difference in peak separation between the \textbf{x} and \textbf{y} scan is due to mesh spacing inhomogeneity.

\begin{figure}\includegraphics[width=8cm]{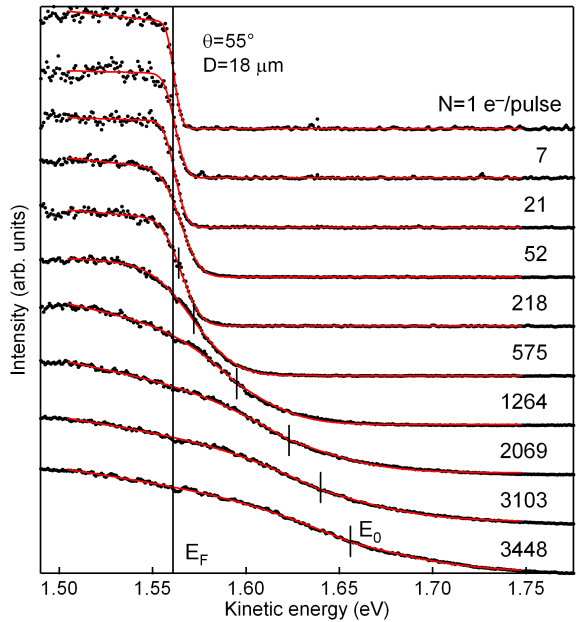}
\caption{\label{fig:3}(Color online) Fermi edge of the Au sample as a function of the number of photo-electrons per pulse ($N$). The spectra are normalized and shifted vertically. The solid lines are fits.}\end{figure}

Fig.\ 3 shows typical integrated photoemission spectra from the gold sample at various laser powers showing the Fermi level broadening and shift. The spot size diameter was 18 $\mu$m. 
The edges of these spectra have been fit to a function of the form:
\begin{equation}
f(E)=a+b\cdot \min(E-E_0,0)+c \cdot \textrm{erfc} \left( \frac{1.665(E-E_0)}{\textrm{FWHM}} \right).
\end{equation}
The parameters $a,b,c,E_0,$ and $\textrm{FWHM}$ are fitting parameters and $E$ is the electron kinetic energy. The first two terms account for the background, which includes an offset and a linear contribution for $E<E_0$. The third term is the complementary error function $\textrm{erfc(x)}$,
\begin{equation}
\textrm{erfc}(x)\equiv1-\textrm{erf}(x)=1-\frac{2}{\sqrt{\pi}}\int^x_0{e^{-t^2}}dt,
\end{equation}
which is commonly used to fit Fermi-Dirac distributions convolved with experimental resolution. The characteristic FWHM is defined to be the full-width-half-maximum value of the associated normalized Gaussian. 

The top of the panel shows a sharp Fermi edge at low power with a FWHM of 7.3 meV, consistent with the experimental energy resolution (4 meV from the analyzer and 4-5 meV from the laser bandwidth) and the sample temperature (about 15 K). The total photo-electron yield measurement indicated an average of 1 electrons per pulse. As the laser power is increased, a greater number of photo-electrons ($N$) are emitted per pulse. This results in a stronger space charge effect, and the edge becomes broader and shifts to higher kinetic energy. 

The shift to higher kinetic energy is due to the fact that the electrons at the Fermi energy have the highest initial kinetic energy, and therefore the Coulomb repulsion of slower photo-electrons can only push them towards higher kinetic energy (neglecting any mirror charge effects). We note that within the laser power range used here, the $\textrm{erfc(x)}$ fit function captures the experimental line shape very well. This was not the case at higher laser power (not shown). 

\section{Experimental results}

\begin{figure}[h] \includegraphics[width=7.5cm]{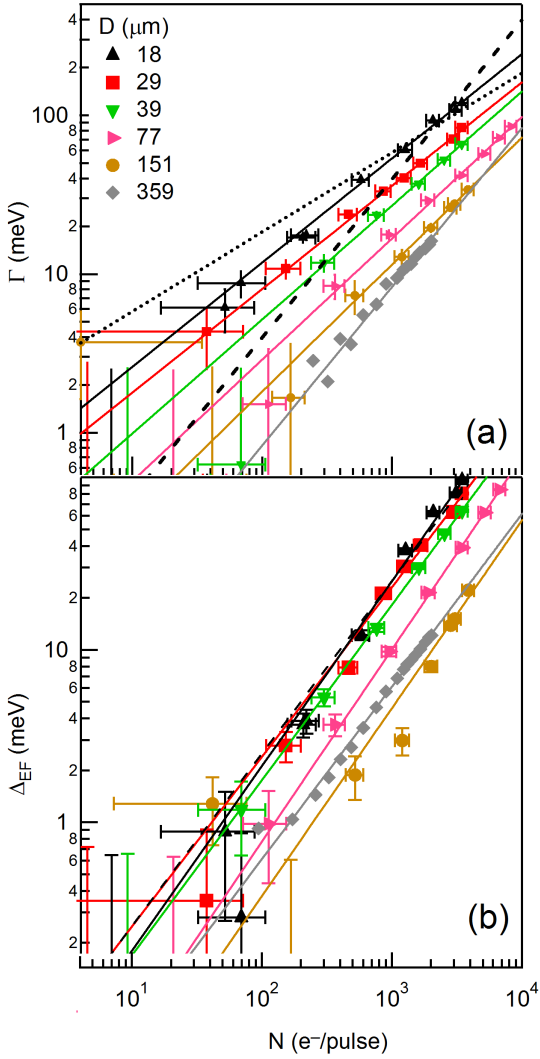}
\caption{\label{fig:4}(Color online) Space charge effect as a function of the number of electrons per pulse ($N$) for various spot diameters ($D$). (a) Broadening ($\Gamma$) of the Fermi edge as a function of $D$ and $N$. The dashed line shows a linear fit and the dotted line a $\sqrt{N}$ fit to the triangular markers. (b) Shift of the Fermi edge ($\Delta_{EF}$) as a function of $N$ for various values of $D$. The dashed lines show a linear fit of the triangular markers. The solid lines are fits to a power law. The data set with diamond shaped markers was obtained at the ALS with 35eV photon energy and was adapted from Ref. \onlinecite{Zhou05b}\@. The y-axis error bars include the errors on the fit and the instrumental resolutions. The x-axis error bar is estimated from the sample current measurement resolution. The spectra are averaged over $\approx$18\degree\ along the analyzer slit with $\theta=55$\degree.} \end{figure}

In Fig.\ 4 we report the shift of the Fermi edge $\Delta_{EF}$, and its broadening $\Gamma$, as a function of number of electrons per pulse ($N$) for several values of the spot size diameter ($D$).  Here $\Delta_{EF}$ is defined by the difference between the measured edge position ($E_0$) and the Fermi energy ($E_F$) while $\Gamma$ is defined as the edge FWHM deconvolved from the sample temperature and experimental resolution. The $\Gamma$ error bars are mainly estimated from fit standard deviation and temperature measurement uncertainty (estimated to be about 3 K). The $\Delta_{EF}$ error bars are the standard deviations plus 0.5 meV\@.
The error on $N$ was estimated to be $\pm(30+10\%)$.

We observe that $\Gamma$ and $\Delta_{EF}$ increase with increasing $N$ and that, as expected, this effect becomes more severe as the spot size decreases. The data from Ref. \onlinecite{Zhou05b} taken in a different regime ($h\nu = 35$ eV and pulse length $=60$ ps),  corresponding to the largest $D$ value shown in the figure (see diamond markers), fall in line with this trend.   
To determine the relation between $\Gamma$, $\Delta_{EF}$, and $N$ we fit the data with the smallest $D$ value using the same power law as the one reported in previous studies.\cite{Zhou05b,Passlack06}  It is clear that neither the linear (see dashed line) nor the $\sqrt{N}$ dependence (dotted line) can describe effectively the data here reported in the ps regime.  To model the transition from the linear to the $\sqrt{N}$ regime we fit the data of Fig.\ 4 with generalized exponents:
\begin{equation}
\Gamma(N) = \alpha_{_\Gamma}\times N^{\beta_{\Gamma}},
\label{Eq:G1}
\end{equation}
\begin{equation}
\Delta_{EF}(N) = \alpha_{_{\Delta EF}}\times N^{\beta_{\Delta EF}}.
\end{equation}

\begin{figure}[h] \includegraphics[width=7.5cm]{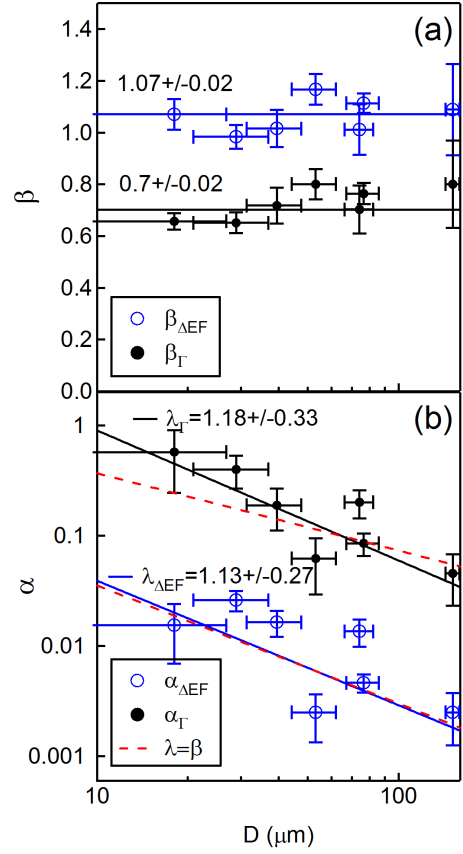}
\caption{\label{fig:5}(Color online) Plot of the the fitting coefficients $\alpha$ and $\beta$ as a function of $D$. The solid lines in panel (a) show the average value of $\beta_{\Gamma}$ and $\beta_{\Delta EF}$. The solid and dashed lines in panel (b) are  fits to a power law.}\end{figure}

In Figs.\ 5(a) and (b), we report the dependence of the fit coefficients $\alpha$ and $\beta$ on $D$, respectively. Remarkably, $\beta$ is essentially independent of $D$ for both $\Gamma$ and $\Delta_{EF}$. More specifically, for the broadening we found $\beta_\Gamma=0.7$, which is an intermediate value between the value (1) reported for synchrotron pulses \cite{Zhou05b} and the value (0.5) reported for fs pulses.\cite{Passlack06}  For the Fermi edge shift we found $\beta_{\Delta EF}=1.1$. In both cases the prefactor $\alpha$ increases with decreasing $D$.
To characterize the space charge effect dependence on $D$ more quantitatively, we fit $\alpha$ with a function of the form 
\begin{equation}
\alpha(D)=\frac{C}{D^\lambda}\; , 
\end{equation}
where $C$ and $\lambda$ are fitting parameters. In a log-log plot, $\alpha(D)$ is a straight line with slope $\lambda$. By substituting $\alpha(D)$ into Eq. (\ref{Eq:G1}), we find that the space charge effect dependence on the number of photo-electrons $N$ and spot diameter $D$ is of the from:
\begin{equation}
f(N,D)=C\times\frac{N^\beta}{D^{\lambda}} \; .
\end{equation}
More specifically we found for the Fermi edge shift: 
\begin{equation}
\Delta_{EF}(N,D)=0.5\times\frac{N^{1.1}}{D^{1.1}} \; ,
\end{equation}
and for the Fermi edge broadening:
\begin{equation}
\Gamma(N,D)=13\times\frac{N^{0.7}}{D^{1.2}}\; ,
\end{equation}
where as in all following formulas $\Delta_{EF}(N,D)$ and $\Gamma(N,D)$ are in meV and $D$ is in $\mu$m.

\begin{figure}[h] \includegraphics[width=7.5cm]{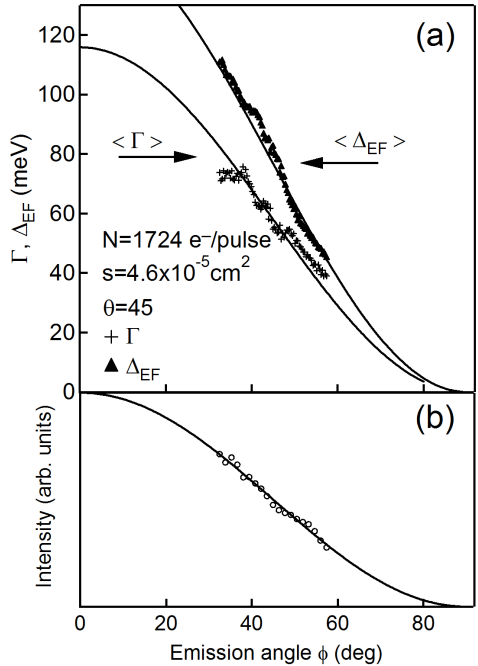}
\caption{\label{fig:6} (a) $\Gamma$ and $\Delta_{EF}$ as a function of the emission angle $\phi$. $\langle\Delta_{EF}\rangle=77$ meV and $\langle\Gamma\rangle=79$ meV are estimated from the fit of the angle-integrated spectrum. (b) Photoemission intensity as a function of $\phi$. The solid lines are fits with the function $h(\phi)= \textrm{const}\times\cos^2(\phi)$.}\end{figure}

To quantify the angular dependence, the angle-resolved results are shown in Fig.\ 6. The emission angle ($\phi$) is along the same axis as $\theta$ with $\phi=0$\degree\ corresponding to normal emission. The angular acceptance of the analyzer was $\approx$ 25\degree\ along the entrance slit. Panel (a) shows the strong dependence of $\Delta_{EF}$ and $\Gamma$ from $\phi$. We show in panel (b) the photoemission intensity measured at $\approx 200$ meV below the Fermi energy. As a guide for the eye, we fit $\Delta_{EF}(\phi)$, $\Gamma(\phi)$ and the photoemission intensity with empirical functions of the form $h(\phi)=\textrm{const}\times\cos^2 \phi$. An offset is added in the case of the photoemission intensity.
 
The strong correlation between the Fermi edge broadening and the photoemission intensity suggests that the angular dependence of the space charge effect is mainly due to the photo-electron emission angular distribution. The effect of the angular integration over the analyzer acceptance angle is different for $\Delta_{EF}$ and $\Gamma$. While the shift of the angle-integrated spectrum ($\langle\Delta_{EF}\rangle=77$ meV) is close to the average value of $\Delta_{EF}(\phi)$ over $\phi$, the broadening of the angle-integrated spectrum ($\langle\Gamma\rangle=79$ meV) is significantly larger than the average value of $\Gamma(\phi)$. This shows that the broadening due to the angular dependence of $\Delta_{EF}(\phi)$ can be larger than just the kinetic broadening. This angular distribution of the space charge effect has a direct effect on the data reported here since the spectra were integrated over $\approx 18$\degree\ with $\theta$=55\degree.

Since we reported the space charge effects as a function of $N$, measurements at different emission angles will yield different values for the coefficient $\alpha$. This should result in a different value of $C$. The exact contribution of the angle integration window and emission angle on the exponents $\beta$ and $\lambda$ is more difficult to estimate.

\section{Simulation results}

The numerical simulations are based on a representative electron energy distribution with a sharp Fermi edge at 1.56 eV\@. The pulse duration was assumed to be rectangular with a duration of 6 ps. The simulations were repeated at least 10 times for each parameters setting. $\Delta_{EF}$ and $\Gamma$ are extracted with the same edge fit function with the exception of the parameters $a$ and $b$ being set to 0. The electron emission was assumed to follow a $\cos^2(\phi)$ distribution following the experimental results of Fig.\ 6 and only electrons with an emission angle between 46\degree\ and 64\degree are integrated in order to mimic the experimental condition. The number of electrons per pulse $N$ varied between 200 and 4000. Spot diameters between 18 $\mu$m and 359 $\mu$m were used. 

\begin{figure}[h] \includegraphics[width=7.5cm]{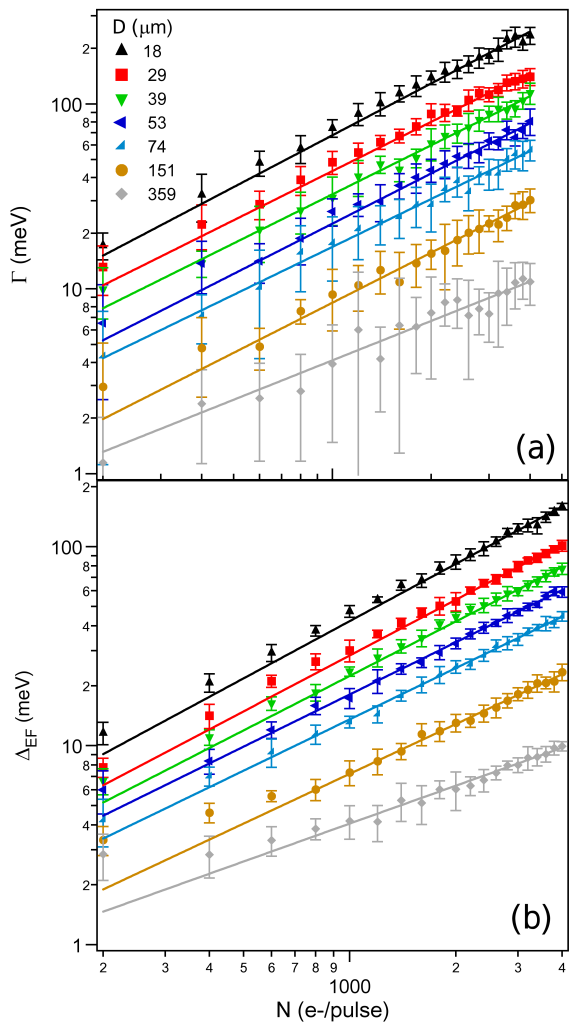}
\caption{\label{fig:7}(Color online)  Simulation of the space charge effect. (a) $\Gamma$ and (b) $\Delta_{EF}$ as a function of $N$ for various $D$ value. The error bars result from multiple simulation runs and represent the standard deviations. The solid lines are fits to a power law.}\end{figure}

\begin{figure}[h] \includegraphics[width=7.5cm]{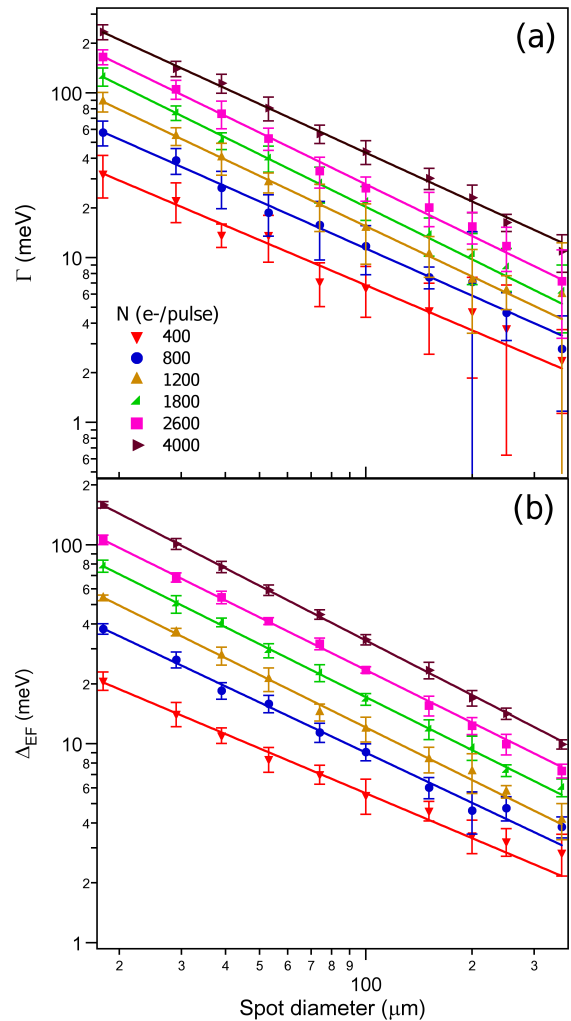}
\caption{\label{fig:8}(Color online)  Simulation of the space charge effect. (a) $\Gamma$ and (b) $\Delta_{EF}$ as a function of $D$ for various $N$ values. The error bars result from multiple simulation runs and represent the standard deviations. The solid lines are fits to a power law.}\end{figure}

Fig.\ 7 shows the simulated broadening $\Gamma$ and shift $\Delta_{EF}$ of the Fermi edge as a function of the number of electrons per pulse $N$ for different spot diameters $D$. Fig.\ 8 shows $\Gamma$ and $\Delta_{EF}$ as a function of the spot diameter $D$ for different numbers of electrons per pulse $N$. We fit the results for $\Gamma$ and $\Delta_{EF}$ with the same power law as in the experimental case and show the comparative results in Table 1. The exponent values are the average value of all the fitting results (not shown). 

\renewcommand\arraystretch{2}
\begin{table}[h]
\centering
\begin{tabular}{r|c|c}
& \hspace{2mm} Simulation \hspace{2mm} & \hspace{2mm} Experiment \hspace{2mm} \\
\hline
$\Delta_{EF}(N,D)=$\hspace{2mm}& $3.1\times$ \large $\frac{N^{0.79}}{D^{0.85}}$&$0.5\times$ \large $\frac{N^{1.1}}{D^{1.1}}$ \\
\hline
$\Gamma(N,D)=$ & $5.5\times$ \large $\frac{N^{0.83}}{D^{1.02}} $  & $13\times$ \large $\frac{N^{0.7}}{D^{1.2}} $ \\
\end{tabular}
\caption{Summary of the measured and simulated exponent values.}
\end{table}
The broadening and the shift of the Fermi edge show a good overall agreement between simulation and experiment. Small systematic deviations are observed in the small effect regime. In particular for $N<600$ and/or $D>200$ $\mu$m.

\section{Discussion}

In the case of the disk-shaped photo-electron cloud produced by a fs pulse, the broadening due to the space charge effect was proposed to be of the form\cite{Passlack06}
\begin{equation}
\Gamma \propto \sqrt{\frac{N}{D}}\; . 
\end{equation}
On the other hand, for the 60 ps synchrotron pulse, the broadening follows a linear relation with $N$. The $\Gamma \propto N^{0.7}$ ($N^{0.8}$ for the simulation) dependence we report for this intermediate regime of laser-based ARPES with 6 ps pulses is between these two values. However, these three regimes cannot be solely compared on a pulse duration basis. The fs pulse analysis is based on a surface state peak (86 meV FWHM) measured at the band minimum in the Brillouin zone center. In this case, the majority of the photo-electrons have a higher kinetic energy. This will affect the space charge effect in two ways: i) the peak position will be shifted to lower kinetic energy; ii) the momentum resolution will decrease resulting in a broader spectrum shifted towards higher kinetic energy. While the total shift in energy ends up being negligible, the contribution of the momentum broadening on $\Gamma$ is not.\cite{Passlack06} This makes a direct comparison difficult but further comparative analysis can be done with our model.

According to the simulation results, the nonlinearity  $\Gamma \propto N^{\beta_\Gamma}$ with $\beta_\Gamma\neq1$ seems to depend largely on the initial photo-electron distribution. More specifically, the simulations show that the exponent $\beta_\Gamma$ decreases as the maximum electron kinetic energy decreases. This exponent is around 1 for high kinetic energies, decreases to $\beta_\Gamma=0.7-0.8$ in the intermediate regime presented here with a maximum electron kinetic energy of about 1.5 eV, and decreases down to $\beta_\Gamma=0.5$ for the fs pulse experiment where the maximum electron kinetic energy was around 1.1 eV.\cite{Passlack06} We found also that the simulation shows no indications that the pulse durations affect this behavior of the exponent  $\beta_\Gamma$.

Similarly, we find the broadening dependence on spot size to be proportional to $D^{-1.2}$ ($D^{-1}$ for the simulation), compared to $D^{-0.5}$ the fs pulse data, though more experimental data are required here to validate the trend. 

It is interesting to note also that the ratio of the two coefficients $\beta_\Gamma$ and $\lambda_\Gamma$ is not 1 as for the fs pulse case, but closer to 0.6-0.8. Since the error bars on $\lambda$ are relatively large, fits with $\lambda=\beta$ are shown with a dashed line in Fig.\ 5 (b). From this fit it is clear that additional data points are needed to conclude on the correct value of the ratio between $\lambda$ and $\beta$. Nevertheless, these results set already a limit in the range in which the analytical model proposed in Ref. \onlinecite{Passlack06} is valid. 
 
\section{Conclusions}
In conclusion, we report the measurement of the space charge effect in the few-ps laser pulse regime. We observe a broadening of the spectral feature exceeding 10 meV and an energy shift exceeding 2 meV with pulses of as few as a 100 photo-electrons for small spot sizes ($\approx$ 20 $\mu$m diameter). For larger spot sizes ($\approx$ 165 $\mu$m diameter) similar effects are observed only as the number of photo-electrons per pulse reaches 600 to 800. We found that neither the linear model that fits synchrotron data nor an $\sqrt{N}$ model that fits fs pulses describe our data effectively, and we introduced a simple phenomenological model with generalized exponents that characterizes the broadening and energy shift of the spectra as a function of spot size and number of photo-electrons over the entire range probed. The results were used as a benchmark for an $N$-body numerical simulation with a good overall agreement. In turn the simulation allowed us to explain the difference observed in different ARPES regimes in terms of initial electron kinetic energy. These results represent a reference for the design and interpretation of photoemission experiments with next-generation light sources, such as free-electron lasers and table-top laser sources.

\begin{acknowledgments}
The laser-ARPES measurements, data analysis and the laser-ARPES chamber setup were supported by the Director, Office of Science, Office of Basic Energy Sciences, Materials Sciences and Engineering Division, of the U.S. Department of Energy
under Contract No. DE-AC02-05CH11231. J.G. was supported in part by a MURI program of the Air Force Office of
Scientific Research, Grant No. FA9550-04-1-0242, which partially contributed to the realization of the Laser-ARPES setup. 
\end{acknowledgments}


\begin{thebibliography}{10}

\bibitem{Clauberg89}
R.~Clauberg, A.~Blacha, {\it J. Appl. Phys.\/} {\bf 65}, 4095 (1989).

\bibitem{Farkas90}
G.~Farkas, C.~T\'oth, {\it Phys. Rev. A\/} {\bf 41}, 4123 (1990).

\bibitem{Gilton90}
T.~L. Gilton, J.~P. Cowin, G.~D. Kubiak, A.~V. Hamza, {\it J. Appl. Phys.\/}
  {\bf 68}, 4802 (1990).

\bibitem{Girardeau91}
C.~Girardeau-Montaut, J.~P. Girardeau-Montaut, {\it Phys. Rev. A\/} {\bf 44},
  1409 (1991).

\bibitem{Aeschlimann95}
M.~Aeschlimann, {\it et~al.\/}, {\it Rev. Sci. Instrum.\/} {\bf 66}, 1000
  (1995).

\bibitem{Qian02}
B.-L. Qian, H.~E. Elsayed-Ali, {\it J. Appl. Phys.\/} {\bf 91}, 462 (2002).

\bibitem{Hellmann09}
S.~Hellmann, K.~Rossnagel, M.~Marczynski-B\"uhlow, L.~Kipp, {\it Phys. Rev.
  B\/} {\bf 79}, 035402 (2009).

\bibitem{Passlack06}
S.~Passlack, {\it et~al.\/}, {\it J. Appl. Phys.\/} {\bf 100}, 024912 (2006).

\bibitem{Zhou05b}
X.~Zhou, {\it et~al.\/}, {\it J. Electron. Spectrosc. Relat. Phenom.\/} {\bf
  142}, 27 (2005).

\bibitem{Liu08}
G.~Liu, {\it et~al.\/}, {\it Rev. Sci. Instrum.\/} {\bf 79}, 023105 (2008).

\bibitem{Hufner}
S.~H\"ufner, {\it Photoelectron Spectroscopy\/} (Springer, Berlin, 1995).

\end{thebibliography}
\end{document}